\documentclass[aps,prl,twocolumn,superscriptaddress,showpacs,showkeys]{revtex4}
\usepackage{epsf}
\usepackage{graphicx}    

\usepackage[T1]{fontenc}


\begin{document}

\def\deeps{\mbox{D($e,e^\prime p_s$)}}
\def\f2neff{$F_{2n}^{\mbox{\tiny eff}}$}
\hyphenation{RCSLACPOL}


\newcommand*{\ANL}{Argonne National Laboratory,  Argonne, Illinois 60439}
\affiliation{\ANL}
\newcommand*{\ASU}{Arizona State University, Tempe, Arizona 85287-1504}
\affiliation{\ASU}
\newcommand*{\UCLA}{University of California at Los Angeles, Los Angeles, California  90095-1547}
\affiliation{\UCLA}
\newcommand*{\CSU}{California State University, Dominguez Hills, Carson, CA 90747}
\affiliation{\CSU}
\newcommand*{\CMU}{Carnegie Mellon University, Pittsburgh, Pennsylvania 15213}
\affiliation{\CMU}
\newcommand*{\CUA}{Catholic University of America, Washington, D.C. 20064}
\affiliation{\CUA}
\newcommand*{\SACLAY}{CEA-Saclay, Service de Physique Nucl\'eaire, F91191 Gif-sur-Yvette, France}
\affiliation{\SACLAY}
\newcommand*{\CNU}{Christopher Newport University, Newport News, Virginia 23606}
\affiliation{\CNU}
\newcommand*{\UCONN}{University of Connecticut, Storrs, Connecticut 06269}
\affiliation{\UCONN}
\newcommand*{\ECOSSEE}{Edinburgh University, Edinburgh EH9 3JZ, United Kingdom}
\affiliation{\ECOSSEE}
\newcommand*{\FIU}{Florida International University, Miami, Florida 33199}
\affiliation{\FIU}
\newcommand*{\FSU}{Florida State University, Tallahassee, Florida 32306}
\affiliation{\FSU}
\newcommand*{\GWU}{The George Washington University, Washington, DC 20052}
\affiliation{\GWU}
\newcommand*{\ECOSSEG}{University of Glasgow, Glasgow G12 8QQ, United Kingdom}
\affiliation{\ECOSSEG}
\newcommand*{\ISU}{Idaho State University, Pocatello, Idaho 83209}
\affiliation{\ISU}
\newcommand*{\INFNFR}{INFN, Laboratori Nazionali di Frascati, 00044 Frascati, Italy}
\affiliation{\INFNFR}
\newcommand*{\INFNGE}{INFN, Sezione di Genova, 16146 Genova, Italy}
\affiliation{\INFNGE}
\newcommand*{\ORSAY}{Institut de Physique Nucleaire ORSAY, Orsay, France}
\affiliation{\ORSAY}
\newcommand*{\BONN}{Institute f\"{u}r Strahlen und Kernphysik, Universit\"{a}t Bonn, Germany}
\affiliation{\BONN}
\newcommand*{\ITEP}{Institute of Theoretical and Experimental Physics, Moscow, 117259, Russia}
\affiliation{\ITEP}
\newcommand*{\JMU}{James Madison University, Harrisonburg, Virginia 22807}
\affiliation{\JMU}
\newcommand*{\KYUNGPOOK}{Kyungpook National University, Daegu 702-701, South Korea}
\affiliation{\KYUNGPOOK}
\newcommand*{\MIT}{Massachusetts Institute of Technology, Cambridge, Massachusetts  02139-4307}
\affiliation{\MIT}
\newcommand*{\UMASS}{University of Massachusetts, Amherst, Massachusetts  01003}
\affiliation{\UMASS}
\newcommand*{\MOSCOW}{Moscow State University, General Nuclear Physics Institute, 119899 Moscow, Russia}
\affiliation{\MOSCOW}
\newcommand*{\UNH}{University of New Hampshire, Durham, New Hampshire 03824-3568}
\affiliation{\UNH}
\newcommand*{\NSU}{Norfolk State University, Norfolk, Virginia 23504}
\affiliation{\NSU}
\newcommand*{\OHIOU}{Ohio University, Athens, Ohio  45701}
\affiliation{\OHIOU}
\newcommand*{\ODU}{Old Dominion University, Norfolk, Virginia 23529}
\affiliation{\ODU}
\newcommand*{\PITT}{University of Pittsburgh, Pittsburgh, Pennsylvania 15260}
\affiliation{\PITT}
\newcommand*{\RPI}{Rensselaer Polytechnic Institute, Troy, New York 12180-3590}
\affiliation{\RPI}
\newcommand*{\RICE}{Rice University, Houston, Texas 77005-1892}
\affiliation{\RICE}
\newcommand*{\Turkey}{Sakarya University, Sakarya, Turkey}
\affiliation{\Turkey}
\newcommand*{\URICH}{University of Richmond, Richmond, Virginia 23173}
\affiliation{\URICH}
\newcommand*{\SCAROLINA}{University of South Carolina, Columbia, South Carolina 29208}
\affiliation{\SCAROLINA}
\newcommand*{\JLAB}{Thomas Jefferson National Accelerator Facility, Newport News, Virginia 23606}
\affiliation{\JLAB}
\newcommand*{\UNIONC}{Union College, Schenectady, NY 12308}
\affiliation{\UNIONC}
\newcommand*{\VT}{Virginia Polytechnic Institute and State University, Blacksburg, Virginia   24061-0435}
\affiliation{\VT}
\newcommand*{\VIRGINIA}{University of Virginia, Charlottesville, Virginia 22901}
\affiliation{\VIRGINIA}
\newcommand*{\WM}{College of William and Mary, Williamsburg, Virginia 23187-8795}
\affiliation{\WM}
\newcommand*{\YEREVAN}{Yerevan Physics Institute, 375036 Yerevan, Armenia}
\affiliation{\YEREVAN}
\newcommand*{\NOWJLAB}{Thomas Jefferson National Accelerator Facility, 
Newport News, Virginia 23606}
\newcommand*{\NOWOHIOU}{Ohio University, Athens, Ohio  45701}
\newcommand*{\NOWUNH}{University of New Hampshire, Durham, New Hampshire 03824-3568}
\newcommand*{\NOWUMASS}{University of Massachusetts, Amherst, Massachusetts  01003}
\newcommand*{\NOWMOSCOW}{Moscow State University, General Nuclear Physics Institute, 119899 Moscow, Russia}
\newcommand*{\NOWMIT}{Massachusetts Institute of Technology, Cambridge, Massachusetts  02139-4307}
\newcommand*{\NOWURICH}{University of Richmond, Richmond, Virginia 23173}
\newcommand*{\NOWODU}{Old Dominion University, Norfolk, Virginia 23529}
\newcommand*{\NOWCUA}{Catholic University of America, Washington, D.C. 20064}
\newcommand*{\NOWGEISSEN}{Physikalisches Institut der Universit\"{a}t Giessen, 35392 Giessen, Germany}
\newcommand*{\NOWLANL}{Los Alamos National Laboratory, Los Alamos, New Mexico 87545}

\title{Measurement of the $x$- and $Q^2$-Dependence of the 
Asymmetry $A_1$ on the Nucleon}


\author {K.V.~Dharmawardane} 
\altaffiliation[Current address: ]{\NOWJLAB}
\affiliation{\ODU}
\author {S.E.~Kuhn} 
     \email{skuhn@odu.edu}
     \thanks{Corresponding author.}
\affiliation{\ODU}
\author{P.~Bosted}
\affiliation{\JLAB}
\author {Y.~Prok} 
\altaffiliation[Current address: ]{\NOWMIT}
\affiliation{\VIRGINIA}
\affiliation{\JLAB}
\author {G.~Adams} 
\affiliation{\RPI}
\author {P.~Ambrozewicz} 
\affiliation{\FIU}
\author {M.~Anghinolfi} 
\affiliation{\INFNGE}
\author {G.~Asryan} 
\affiliation{\YEREVAN}
\author {H.~Avakian} 
\affiliation{\INFNFR}
\affiliation{\JLAB}
\author {H.~Bagdasaryan} 
\affiliation{\YEREVAN}
\affiliation{\ODU}
% ...
\collaboration{The CLAS Collaboration}
     \noaffiliation


\date{\today}

\begin{abstract}
We report results for the virtual photon 
asymmetry $A_1$ on the nucleon from 
new Jefferson Lab measurements. The experiment, which used the CEBAF Large Acceptance Spectrometer and
longitudinally polarized proton ($^{15}$NH$_3$) and deuteron ($^{15}$ND$_3$) targets, 
collected data with a longitudinally polarized
electron beam at energies between 1.6 GeV and  5.7 GeV. In the present paper, 
we concentrate on our results
for $A_1(x,Q^2)$ and the related ratio $g_1/F_1(x,Q^2)$
in the resonance and the deep inelastic regions for our lowest and highest beam
energies, covering a range in momentum transfer $Q^2$ from 0.05 to 5.0 GeV$^2$ and 
in final-state invariant mass $W$ up
to about 3 GeV.
Our data show detailed structure in the resonance region, which leads to a strong $Q^2$--dependence
of $A_1(x,Q^2)$ for $W$ below 2 GeV. At higher $W$, a smooth
approach to the scaling limit, established by earlier experiments, can be seen, but $A_1(x,Q^2)$ is
not strictly $Q^2$--independent. We add significantly to the world data set at high $x$, up to
$x = 0.6$. 
Our data exceed the SU(6)-symmetric quark model expectation for both the proton and the 
deuteron while being consistent with a negative $d$-quark polarization up to our highest $x$. 
This data set
should improve next-to-leading order (NLO) pQCD fits of the parton polarization distributions.
\end{abstract}
\keywords{Spin structure functions, nucleon structure}
\pacs{13.60.Hb, 13.88.+e , 14.20.Dh}
\maketitle

%
The spin structure of the nucleon has been investigated in a series of much-discussed
polarized lepton scattering
experiments over the last 25 years 
\cite{E80final,E130g1p,EMCfinal,SMCfinal,E142Long,E143Long,E154g1,E155Q2,HERMESfinal,HallAMoments,HallA_largex,EG1a_p,EG1a_d}. 
These measurements, most of which covered the deep
inelastic scattering (DIS) region of large final-state invariant mass $W$ and momentum transfer $Q^2$,  
compared the $Q^2$-dependence of the polarized structure function $g_1$ with QCD expectations
and shed new light on the structure of the nucleon. 
Among the most surprising results was the realization
that only a small fraction of the nucleon spin (20\% -- 30\%) is carried by the quark helicities, 
in disagreement
with quark model expectations of 60\% -- 75\%.
This reduction is often attributed to the effect
of a negatively polarized quark sea at low momentum fraction $x$, which is typically not included in
quark models (see the paper by Isgur~\cite{SpinQMIsgur} 
for a detailed discussion).

For a more complete understanding of the quark structure of the nucleon, 
it is advantageous to concentrate on a kinematic region where the
scattering is most likely to occur
from a valence quark in the nucleon carrying more than a fraction $x = 1/3$ of the
nucleon momentum. In particular, the virtual photon asymmetry, $A_1(x) \approx g_1(x)/F_1(x)$,
(where $F_1$ is the usual unpolarized structure function) can be 
(approximately) interpreted in terms
of the polarization $\Delta u/u$ and $\Delta d/d$ of the valence $u$ and $d$ quarks in the proton
in this kinematic region, where the contribution from sea quarks is minimized. This asymmetry also
has the advantage of showing smaller scaling violations than the structure functions $g_1$ and $F_1$
individually~\cite{E143Long,E155Q2}, 
making a comparison with various theoretical models and predictions more straightforward.

By measuring $A_1(x)$ at large $x$, one can test different predictions about the limit of
$A_1(x)$ as $x \rightarrow 1$. Non-relativistic Constituent Quark Models (CQM) based on SU(6) symmetry
predict $A_1(x) = 5/9$ for the proton, $A_1(x) = 0$ for the neutron and $A_1(x) = 1/3$ for
the deuteron (modified by a factor $(1 - 1.5 w_D)$ for the D-state probability $w_D$ 
in the deuteron wave function). 
Quark models that include some mechanism of SU(6) symmetry breaking 
({\it e.g.}, one-gluon exchange hyperfine interaction between quarks~\cite{SpinQMIsgur}) 
predict that $A_1(x) \rightarrow 1$  for all three targets as 
$x$ tends to 1.
This is because target remnants with total spin 1 are suppressed relative to those with spin 0. 
The same limit for $x \rightarrow 1$ is also predicted by pQCD~\cite{SpinQCD}, 
because hadron helicity conservation suppresses the
contribution from quarks anti--aligned with the nucleon spin. In this case, 
$A_1(x)$ would be predicted to be more positive at moderately large $x<1$ because 
 both $u$ and $d$ quarks contribute with positive polarization~\cite{BBS}.
Finally, a recent paper~\cite{ClosMeln}
connected the behavior of $A_1(x)$ at large $x$ with the dynamics of resonance
production via duality, leading to several predictions for the approach to
$A_1(x \rightarrow 1)=1$ that depend on the mechanism of SU(6) symmetry breaking. 

Clearly, measurements of the asymmetry $A_1$ at moderate to high $x \ge 0.3$
 are an indispensable tool to
improve our understanding of the valence structure of the nucleon.
Although many data already exist on $A_1(x,Q^2)$, most of the high-energy data
have very limited statistics at large $x$ and therefore large uncertainties; high-precision data so far exist only
for a $^3$He target~\cite{HallA_largex}
(which can be used to approximate $A_1$ for a free neutron). 
Those data show for the first time a positive asymmetry $A_1^n$ at large $x$, but agree better with
predictions~\cite{SpinQMIsgur} that assume negative $d$-quark polarization $\Delta d/d$ even at large $x$. 

In this paper, we report
the first high-precision measurement of $A_1(x,Q^2)$ for the proton and the deuteron at moderate to large $x$ ($x \ge  0.15$)
over a range of momentum transfers $Q^2 =0.05...5.0$ GeV$^2$, 
covering both the resonance and the deep inelastic region.

%
The data described in this paper
were collected during the second polarized target run (2000-2001) with CLAS
 in Hall B of the Thomas Jefferson National Accelerator
Facility (TJNAF -- Jefferson Lab). 
Results from the first run  with beam energies of 4.2 and 2.5 GeV
were recently published~\cite{EG1a_p,EG1a_d}. The present data extend the
kinematic coverage significantly to both lower and higher values of $Q^2$
(covering nearly two orders of magnitude, instead of only one), and 
to higher values of $W$, covering much more of the DIS region
(nearly doubling the range in $x$).
Longitudinally
polarized electrons of several beam energies around 1.6 GeV and 5.7 GeV
were scattered off longitudinally polarized ammonia targets
--- $^{15}$NH$_3$ and $^{15}$ND$_3$ --- and detected in the 
CEBAF Large Acceptance 
Spectrometer (CLAS).
A detailed description of CLAS may be found in
Ref.~\cite{Mecking:2003zu}.  The spectrometer is equipped with a
superconducting toroidal magnet and three drift chamber  regions
that cover up to 80\% of the azimuthal angles and reconstruct 
the momentum of charged particles scattering
within a polar angular range between 8$^\circ$ and 142$^\circ$. 
(Due to obstruction by the polarized target Helmholtz coils only
scattering angles up to 50$^\circ$ were accessible during our experiment.) 
We used both the inbending (for electrons) and the outbending
torus magnetic field orientations, to extend the coverage in $Q^2$.  An
array of scintillator counters
covers the above
angular range and is used to determine the time of flight for
charged particles.
A forward angle electromagnetic calorimeter
16 radiation lengths thick covers  polar angles up to  45$^\circ$ and
is used along with the drift chambers to separate pions from
electrons for this analysis.  A gas Cherenkov detector
covering the same 
angular range as the calorimeter is used in conjunction with
the calorimeter to create a coincidence trigger, and to reject
pions.

The target material was kept in a 1 K liquid Helium bath and
was polarized via Dynamic Nuclear Polarization
(DNP)~\cite{target}. The target polarization was monitored online
using a Nuclear Magnetic Resonance (NMR) system. The beam
polarization was measured at regular intervals with a M\o ller polarimeter.
The product of beam and target polarization
($P_b P_t$) was determined from the well-known asymmetry for
elastic (quasielastic) scattering from polarized protons (deuterons),
measured simultaneously with inelastic scattering.
For the 1.6 GeV data set, the average polarization product was
$P_bP_t=0.54 \pm 0.005$  
($0.18 \pm 0.007$) for the $^{15}$NH$_3$ ($^{15}$ND$_3$) target.
The corresponding value for the 5.7 GeV data set was $0.51 \pm 0.01$ 
($0.19 \pm 0.03)$.



%
The data analysis proceeds along the following steps
(see Ref.~\cite{EG1a_d} for details). We
first extract the raw count rate asymmetry
$A_{||}^{raw} = (N^+ - N^-)/(N^+ + N^-)$,
where the electron count rates for anti-parallel ($N^+$) and parallel ($N^-$) electron
and target polarization are normalized to the (live-time gated)
beam charge for each helicity.  The background due to misidentified pions and
electrons from decays into $e^+ e^-$ pairs has been subtracted from these rates.
We divide the result by the product
of beam and target polarization $P_b P_t$ and correct for the contribution
from non-hydrogen nuclei in the target. For this purpose, we use auxiliary
measurements on $^{12}$C, $^4$He and pure $^{15}$N targets. We then combine the
asymmetries for different beam and target polarization directions, thereby
reducing any systematic errors from false asymmetries (no significant 
differences between the different polarization sets were found).
Finally we apply radiative corrections using the code RCSLACPOL~\cite{E143Long}
which follows the prescription by Kuchto and Shumeiko~\cite{Kukhto}
for the internal corrections and by Tsai~\cite{Tsai}
for the external corrections. The (quasi-)elastic radiative tail contribution to the
denominator of the asymmetry is treated as a further dilution factor $f_{RC}$. 

The final result is the longitudinal
(Born) asymmetry $A_{||} = D (A_1 + \eta A_2)$,
where the depolarization factor 
$D = (1-E'\epsilon /E)/(1+\epsilon R)$, $E$ ($E'$) is
the beam (scattered electron) energy, 
$\epsilon = (2 E E' - Q^2/2)/(E^2+E'^2 + Q^2/2)$ is the virtual photon polarization,
$R \stackrel{<}{\sim} 0.2$
is the ratio of the longitudinal to the transverse
photoabsorption cross section and $\eta
= (\epsilon\sqrt{Q^2})/(E - E'\epsilon)$. 
 $A_2$ is the longitudinal-transverse interference virtual photon asymmetry.
 We use the standard notations for the
energy transfer, $\nu = E - E'$,
 and four-momentum transfer squared,  $Q^2 = 4 E E' \sin^2(\theta/2)$.

\begin{figure}
\epsfxsize=\linewidth
\epsfbox{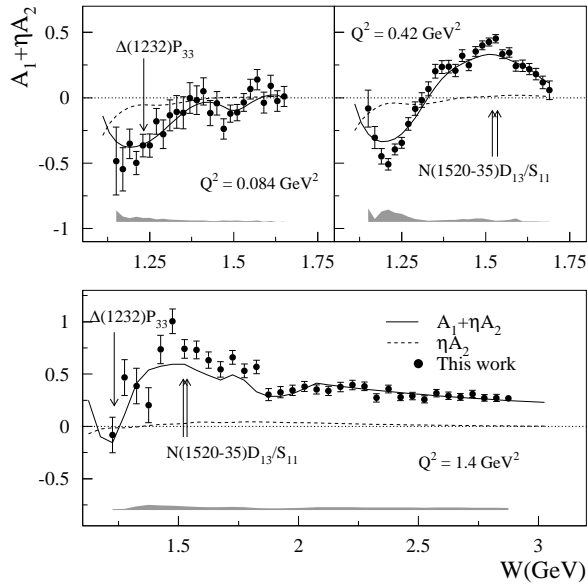}
\caption{Results for the asymmetry $A_{||}/D = A_1 + \eta A_2$ on the proton
versus final-state invariant mass $W$, for three bins in $Q^2$.
  Arrows indicate the masses
of several resonances. The first two panels show data obtained with
1.6 GeV beam energy, while the last panel comes from the 5.7 GeV data.
The solid line close to the data points is the result for $A_{||}/D$ of our parametrization of previous
world data. The dashed line close to zero is the 
estimated contribution from the unmeasured
asymmetry $A_2$ to $A_{||}/D$. Bands at the bottom of all figures indicate
systematic errors.}
\label{fig:A1res}\end{figure}

Finally, using a 
parametrization of the world data~\cite{E143Long,E155Q2}
to model $A_2$ and $R$, we extract $A_1$ and the closely related ratio
$g_1/F_1$:
\begin{equation}
\frac{g_1}{F_1}(x,Q^2)  = \frac{1}{(\gamma^2+1)}\left( \frac{A_{||}}{D}  
+( \gamma - \eta) A_2 \right) 
\label{eq:g1F1}
\end{equation}
with $\gamma^2 = Q^2/\nu^2$.
The extraction of this ratio is typically less dependent on the unmeasured asymmetry, $A_2$,
than that of the asymmetry $A_1$.
Our 
parametrization includes
input from phenomenological models
AO~\cite{AO} and MAID~\cite{MAID} 
as well as fits to the polarized data from the first run with CLAS~\cite{EG1a_p,EG1a_d}
and to unpolarized structure functions measured in Jefferson Lab's Hall C~\cite{Eric}.
More details of the 
parametrization and the data analysis can be found
in Ref.~\cite{EG1a_d}. Since $A_1$ and $g_1/F_1$ are
independent of beam energy for given $(x,Q^2)$ values,
we combine (after consistency checks) our results
for each bin in ($x, Q^2$) for all beam energies and CLAS torus magnetic field settings.

To estimate systematic uncertainties on our final results, we vary all input parameters
and models within realistic limits and study the induced variations of the asymmetry $A_1$.
We then add all these variations in quadrature to get the total systematic uncertainty. Among
the sources of systematic errors we considered are uncertainties on the product of beam
and target polarization and various inputs in our determination of the dilution factor (target
dimensions, nuclear cross sections, and contributions from polarized nuclei other than
the hydrogen isotope under consideration). We also estimate the remaining contribution
from misidentified pions and electrons from pair-symmetric decay processes. Finally,
we varied all model parametrizations for unpolarized ($F_1, R$) and polarized ($A_1, A_2$)
structure functions used both in the extraction of $A_1$ and $g_1/F_1$ and in our
radiative corrections. Systematic errors are indicated by shaded bands in the figures.

%
\begin{figure}
\epsfxsize=\linewidth
\epsfbox{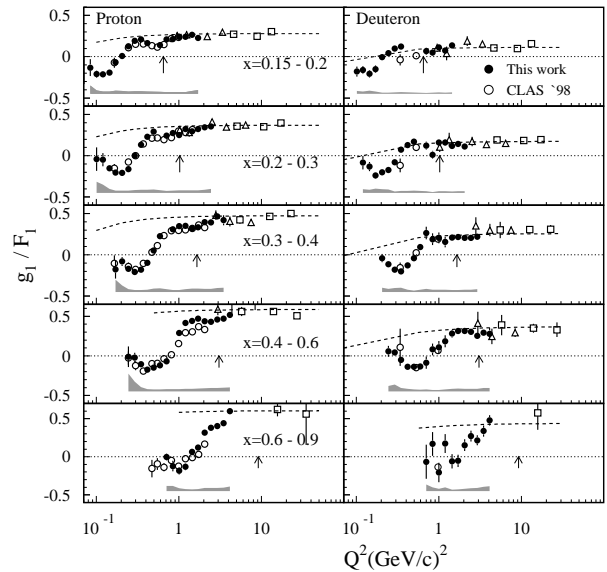}
\caption{Measured ratio $g_1/F_1$ as a function of momentum transfer squared
$Q^2$ for several bins in $x$ for the proton (left) and the deuteron (right).
A few data points
from SLAC experiments E143~\protect{\cite{E143Long}} (open triangles)
 and E155~\protect{\cite{E155Q2}} (open squares)
  are also shown for comparison, as well as data from the
 first run with CLAS~\protect{\cite{EG1a_p,EG1a_d} (open circles)}. 
 The dashed line represents our
parametrization of the world data in the DIS region~\protect{\cite{E155Q2}}. Arrows indicate the
conventional limit of the
resonance region at $W = 2$ GeV.}
\label{fig:g1f1}\end{figure}

A small sample of our results on the asymmetry $A_{||}/D$ for the proton 
is shown in Fig.~\ref{fig:A1res}. Since
the asymmetry $A_2$ contributes only very little to these data
(see dashed line in the figure), they are essentially equal
to $A_1$. A strong dependence of this asymmetry on the final state mass $W$ can be seen,
especially at low $Q^2$ (top left panel). Our total data set covers 19 bins in $Q^2$, with similar
statistics for the deuteron. The entire data set is available at the CLAS Physics 
Database~\cite{CDB} or by request from the authors.
These data can be used to constrain transition amplitudes for resonances of different spin and
isospin which partially overlap with each other and the non-resonant background. For instance,
in the region of the $\Delta(1232)$, the asymmetry is negative at low $Q^2$, since the transition
to the  $\Delta$ is dominated by the $A_{3/2}$ amplitude, while at larger $Q^2$ this amplitude
seems to be suppressed and the non-resonant background becomes more dominant. Similarly,
around $W = 1.53$ GeV, the asymmetry makes a rapid transition from being slightly negative
at small $Q^2$ to
large positive values even at rather moderate $Q^2$, indicating that the $A_{3/2}$ amplitude
for the transition to the $D_{13}$ resonance becomes less important than the $A_{1/2}$
amplitude for the transition to both the $D_{13}$ and $S_{11}$ resonances.

\begin{figure}
\epsfxsize=\linewidth
\epsfbox{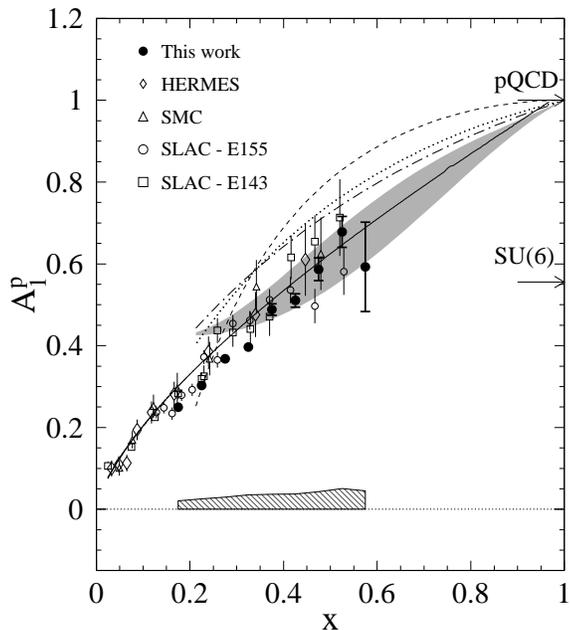}
\caption{Results for the asymmetry $A_1(x)$ on the proton. 
Filled circles show our data in the deep inelastic region ($W > 2$ GeV, $Q^2 > 1$ GeV$^2$)
while the remaining open symbols are for data from several previous 
experiments~\protect{\cite{HERMESfinal,SMCfinal,E155Q2,E143Long}}. 
The SU(6) expectation for all $x$ is indicated by the arrow.
 The solid line shows our parametrization of the world data at a fixed $Q^2 = 10$
GeV$^2$. The shaded band covers a range
of calculations by Isgur\protect{~\cite{SpinQMIsgur}} that model the hyperfine--interaction breaking of 
SU(6) symmetry. The remaining three curves correspond to different scenarios of SU(6) symmetry
 breaking as presented in the paper by Close and Melnitchouk\protect{~\cite{ClosMeln}}:
helicity-1/2 dominance (dashed), spin-1/2 dominance (dotted) and symmetric wave function suppression (dash-dotted).
 }
 \label{fig:A1p}\end{figure}

The closely related ratio of structure functions, $g_1/F_1$, is shown in Fig.~\ref{fig:g1f1}
 as a function of $Q^2$, averaged over several bins in $x$. The new data are in good
 agreement with the results of the first run with CLAS~\cite{EG1a_p,EG1a_d}. In the
DIS region, both $g_1$ and $F_1$ are expected to have only logarithmic
scaling violations, and their ratio has been found to be nearly independent of
 $Q^2$ in previous experiments (see, for example, the SLAC data~\cite{E143Long,E155Q2}
reproduced in Fig.~\ref{fig:g1f1}). Our data show a clear decrease in this asymmetry with decreasing $Q^2$;
 in particular, for the proton they fall below the DIS parametrization around $Q^2=1$ GeV$^2$ 
and small $x$.
This $Q^2$-dependence becomes much more pronounced in the
region of the nucleon resonances (at $Q^2$ below the limits indicated by arrows in Fig.~\ref{fig:g1f1}), 
leading to a strong deviation of the data from
a smooth extrapolation of DIS data~\cite{E155Q2}
(dashed lines in Fig.~\ref{fig:g1f1}).
This is a direct consequence of the fact that $W$ varies with $Q^2$ at fixed $x$ and
reflects the $W$-dependence seen in Fig.~\ref{fig:A1res}.
For kinematics corresponding to the excitation of the $\Delta$ resonance
(at the lowest $Q^2$ in each panel), the asymmetry
is much reduced and even changes sign relative to the DIS region at small $Q^2$ due to
the dominance of the $A_{3/2}$ amplitude.
The data above $W=2$ GeV can be incorporated into NLO fits of spin structure
functions to improve the precision with which polarized parton distribution functions are known.

\begin{figure}
\epsfxsize=\linewidth
\epsfbox{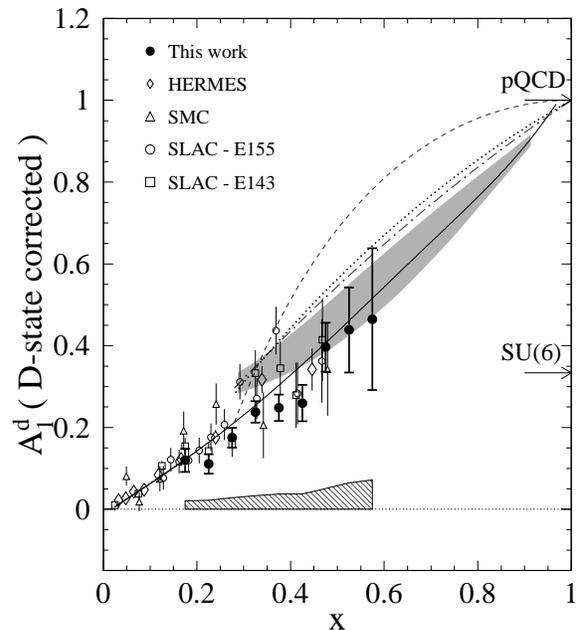}
\caption{
Results for the asymmetry $A_1(x)$ on the
deuteron. The lines and symbols have the same meaning as in Fig.~\protect{\ref{fig:A1p}}.
The data are divided by $(1 - 1.5 w_D) \approx 0.925$ to correct for the deuteron
D-state probability $w_D$, while the model predictions are for an isoscalar (proton plus neutron) target.
}
\label{fig:A1d}\end{figure}

The results for $A_1(x)$, averaged 
over $Q^2 >1$ GeV$^2$ and  $W > 2$ GeV, are shown
in Fig.~\ref{fig:A1p} for the proton and in Fig.~\ref{fig:A1d}
for the deuteron. 
At small $x$, where our average $Q^2$ is close to 1 GeV$^2$,
the data fall below our parametrization of the world data with $Q^2 = 10$ GeV$^2$ (solid line).
This deviation is due to the $Q^2$-dependence shown in
Fig.~\ref{fig:g1f1} (note that $A_1$ and $g_1/F_1$ are very close in this kinematic region). In contrast,
both of our data sets 
exceed the SU(6) limits at $x$ above 0.45.
The hyperfine interaction model of SU(6) symmetry breaking by 
Isgur~\cite{SpinQMIsgur} (grey band in figures) is closest
to the data.
Of the different mechanisms for SU(6) symmetry breaking considered
by Close and Melnitchouk~\cite{ClosMeln}, the model with suppression of the symmetric quark wave function
(dot-dashed curve in Figs. ~\ref{fig:A1p},\ref{fig:A1d})
deviates least from the data. In general, our results are in better agreement with models
(like the two mentioned above) in
which the ratio of down to up quarks, $d/u$, goes to zero and the polarization of down quarks,
$\Delta d/d$ tends to stay negative for rather large values of $x$, in 
contrast to the behavior expected from hadron helicity conservation~\cite{SpinQCD, BBS}.
This is also in agreement with the findings by the experiment on
$^3$He~\cite{HallA_largex} in Jefferson Lab's Hall A.

\begin{figure}
\epsfxsize=\linewidth
\epsfbox{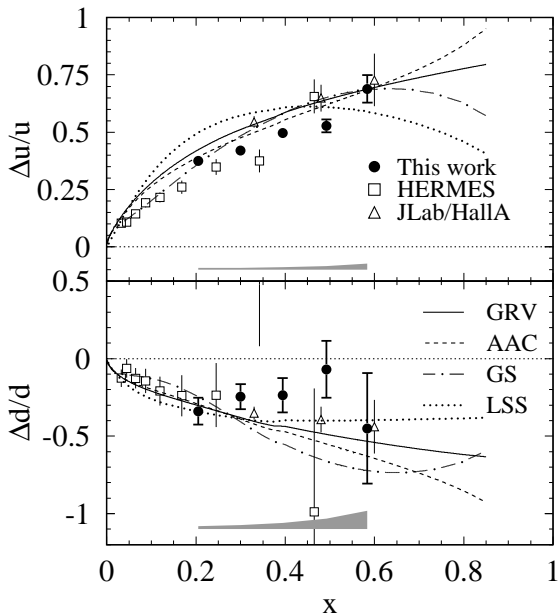}
\caption{Quark polarizations $\Delta u/u$ and $\Delta d/d$
extracted from our data. Included are all data above $W=1.77$ GeV and $Q^2 = 1$ GeV$^2$.
Also shown are semi-inclusive results from Hermes~\protect{\cite{HERMESfinal}} and
inclusive results from Hall A data~\protect{\cite{HallA_largex}} 
combined with previous data from CLAS~\protect{\cite{EG1a_p}}  . 
The solid line is from the NLO fit to
the world data by GRV~\protect{\cite{GRV}}, the dashed line is from the AAC fit~\protect{\cite{AAC}},
the dash-dotted line is from Gehrmann and Stirling~\protect{\cite{Gehrmann}} and the dotted line
indicates the latest fit from LSS~\protect{\cite{LSS}} .}
\label{fig:delq}\end{figure}

Within a naive quark--parton model (and ignoring any contribution from strange quarks), 
we can estimate the quark polarizations
$\Delta u/u$ and $\Delta d/d$ directly from our data by combining the results 
for $g_1$ from
the proton and the deuteron (including some nuclear corrections for the deuteron
D-state and Fermi motion)
with our parametrization of the world data on $F_1^p$ and $F_1^d$:
\begin{eqnarray}
{{\Delta u}\over{u}} & \approx & {{5 g_1^p - 2 g_1^d/(1 - 1.5 w_D)}\over{5 F_1^p - 2 F_1^d}} ; \\
{{\Delta d}\over{d}} & \approx & {{8 g_1^d/(1 - 1.5 w_D) - 5 g_1^p}\over{8 F_1^d - 5 F_1^p}} .
\end{eqnarray} 
The result (Fig.~\ref{fig:delq}) has relatively large statistical errors for $\Delta d/d$,
since neither $A_1^p$ nor $A_1^d$ are very sensitive to $\Delta d/d$. (We included data
down to $W$ = 1.77 GeV in our estimate for the highest $x$ points to reduce those errors
somewhat; at these rather large values of $Q^2 > 3$ GeV$^2$ we expect little deviation
from the DIS limit in this $W$ range).
Our estimate is consistent with the result from the $^3$He experiment~\cite{HallA_largex},
showing no indication of a sign change to positive values up to $x \approx 0.6$. At the same time,
our data for $\Delta u/u$ are the statistically most precise available at this time, and show
a consistent trend towards $\Delta u/u = 1$  at our highest $x$ points.
While the absolute values of $\Delta u/u$ and $\Delta d/d$ might be somewhat different
from more sophisticated NLO DGLAP analyses (like the curves shown in Fig.~\ref{fig:delq}),
the error bars in Fig.~\ref{fig:delq} 
give an indication of the possible improvement in precision when our data are included
in such fits.


%
In summary, we have measured the virtual photon asymmetry $A_1$ and the related
ratio $g_1/F_1$ of structure functions on the proton and the deuteron 
with unprecedented precision, at high $x$ and 
over a large kinematic range in $x$ and $Q^2$. Our data span the resonance
region $W < 2$ GeV and extend into the DIS region. They contribute to our
knowledge of the valence
quark structure of the nucleon and its excited states, and can be used to improve NLO fits
for the extraction of polarized parton distribution functions. Our
data confirm a clear increase in the polarization of valence $u$ quarks
at high $x$ as expected by pQCD and various models of SU(6) symmetry breaking;
on the other hand, the polarization of the $d$ quarks seems to remain negative up to
the highest values of $x$ accessible to our experiment.
Future measurements, in particular with the energy-upgraded Jefferson Lab accelerator,
will be able to extend these data with improved precision to higher values of $x$
(exceeding $x \approx 0.8$),
allowing a definite test of various models of SU(6) symmetry breaking.

\section*{Acknowledgments}
We would like to acknowledge the outstanding efforts of the staff
of the Accelerator and the Physics Divisions at Jefferson Lab that made
this experiment possible.  This work was supported in part by the
Italian Instituto Nazionale di Fisica Nucleare, the French Centre
National de la Recherche Scientifique, the French Commissariat \`{a}
l'Energie Atomique, the U.S. Department of Energy and National
Science Foundation, the Emmy Noether grant from the Deutsche Forschungs
Gemeinschaft and the Korean Science and Engineering Foundation.
The Southeastern Universities Research Association (SURA) operates
the Thomas Jefferson National Accelerator Facility for the
United States Department of Energy under contract DE-AC05-84ER-40150.

\bibliographystyle{apsrev}


\end{document}